\documentclass[10pt,twocolumn,journal]{IEEEtran}

\usepackage{lettrine}
\usepackage{amsmath}
\usepackage{cases}
\usepackage{bm}
\usepackage{txfonts}
\usepackage{amssymb}
\usepackage{cite}
\usepackage{graphicx}
\usepackage{psfrag}
\usepackage{url}
\usepackage{stfloats}
\usepackage{algorithm}
\usepackage{algorithmic}        
\usepackage{multirow}
\usepackage{color}
\usepackage{romannum}
\usepackage{lipsum}
\usepackage{stfloats}
\usepackage[utf8]{inputenc}
\usepackage[skip=2pt,font=scriptsize]{caption}
\usepackage[skip=2pt,font=scriptsize]{subcaption}
\usepackage{tabularx}
\usepackage{booktabs}

\ifCLASSINFOpdf

\else
\fi

\begin{document}
\pagenumbering{arabic}
\title{Synergizing Generative Artificial Intelligence and Beyond Diagonal RIS for Proactive mmWaves Communication}


\author{Abuzar B. M. Adam,  ~\IEEEmembership{Member,~IEEE}, Elhadj Moustapha Diallo,  Mohammed A. M. Elhassan
\thanks{Abuzar B. M. Adam is with the Interdisciplinary Centre for Security, Reliability and Trust (SnT), University of Luxembourg, Luxembourg.}
\thanks{E. M. Diallo is with the School of Communications and Information Engineering, Chongqing University of Posts and Telecommunications, Chongqing, China}
\thanks{Mohammed A. M. Elhassan is with School of Mathematics and Computer Science, Zhejiang Normal University, China.}}
\maketitle

\begin{abstract}
In this work, we explore UAV-assisted reconfigurable intelligent surface (RIS) technology to enhance downlink communications in wireless networks. By integrating RIS on both UAVs and ground infrastructure, we aim to boost network coverage, fairness, and resilience against challenges such as UAV jitter. To maximize the minimum achievable user rate, we formulate a joint optimization problem involving beamforming, phase shifts, and UAV trajectory.
To address this problem, we propose an adaptive soft actor-critic (ASAC) framework. In this approach, agents are built using adaptive sparse transformers with attentive feature refinement (ASTAFER), enabling dynamic feature processing that adapts to real-time network conditions. The ASAC model learns optimal solutions to the coupled subproblems in real time, delivering an end-to-end solution without relying on iterative or relaxation-based methods.
Simulation results demonstrate that our ASAC-based approach achieves better performance compared to the conventional SAC. This makes it a robust, adaptable solution for real-time, fair, and efficient downlink communication in UAV-RIS networks.
\end{abstract}

\section{Introduction}

The sixth generation of wireless networks (6G) is anticipated to tackle key terrestrial network challenges while providing high reliability and massive connectivity \cite{bb1}. Reconfigurable intelligent surfaces (RIS) have recently garnered attention due to their ability to extend network coverage, reduce energy consumption, and enhance signal quality, particularly in scenarios with line-of-sight (LoS) blockages \cite{bb2}. RIS utilizes numerous passive antennas to intelligently reflect signals, thereby improving both coverage and communication quality \cite{bb1}.

Unmanned aerial vehicles (UAVs), which operate without onboard human pilots, have emerged as valuable assets in wireless networks due to their flexibility, mobility, and capability to establish robust communication links compared to fixed terrestrial setups \cite{bb3,bb4,bb5,bb6}. UAVs can further augment network performance by aiding diverse applications such as satellite communications \cite{bb7,bb8}, vehicular networks \cite{bb9}, and visible light communication systems\cite{bb10}. The combined deployment of UAVs and RIS, particularly in flying RIS configurations (UAV-mounted RIS), allows UAVs to function as mobile signal reflectors that dynamically enhance network coverage, especially in challenging environments where connectivity and LoS are limited \cite{bb11}.

In our work, we focus on UAV-RIS-enhanced wireless communications with a specific emphasis on downlink optimization. We aim to maximize minimum user rate in multi-user settings, jointly optimizing power allocation, phase shifts, and UAV trajectory. Prior studies have addressed these optimization challenges for UAV-RIS systems, focusing on key parameters like UAV trajectory and RIS phase shifts to enhance reliability, reduce latency, and improve spectral efficiency \cite{bb12,bb13,bb14}. However, many of these solutions employ traditional methods, such as successive convex approximation (SCA) and semi-definite relaxation (SDR), which are computationally intensive and depend heavily on closed-form solutions. These approaches face limitations in real-time deployment due to their reliance on iterative processes, often using outdated channel state information (CSI) which may degrade system performance in highly dynamic UAV environments \cite{bb15,bb16}.

Given the nonconvexity and dynamic nature of UAV deployment problems, deep reinforcement learning (DRL) has emerged as a powerful alternative. Techniques such as double deep Q-networks (DDQN) and deep deterministic policy gradients (DDPG) have been used to optimize UAV placement and RIS phase shifts in UAV-aided networks, showing potential for handling the complexity of such tasks with reduced computational delay and improved adaptability to environmental uncertainties \cite{bb17,bb18,bb19,bb20}. However, these DRL approaches still face challenges in real-time applications, particularly due to their reliance on extensive training data, complexity in handling large-scale Markov decision processes (MDPs), and potential degradation in performance as problem size increases \cite{bb21}.
\begin{figure*}[htbp]
\centerline{\includegraphics[width=5in]{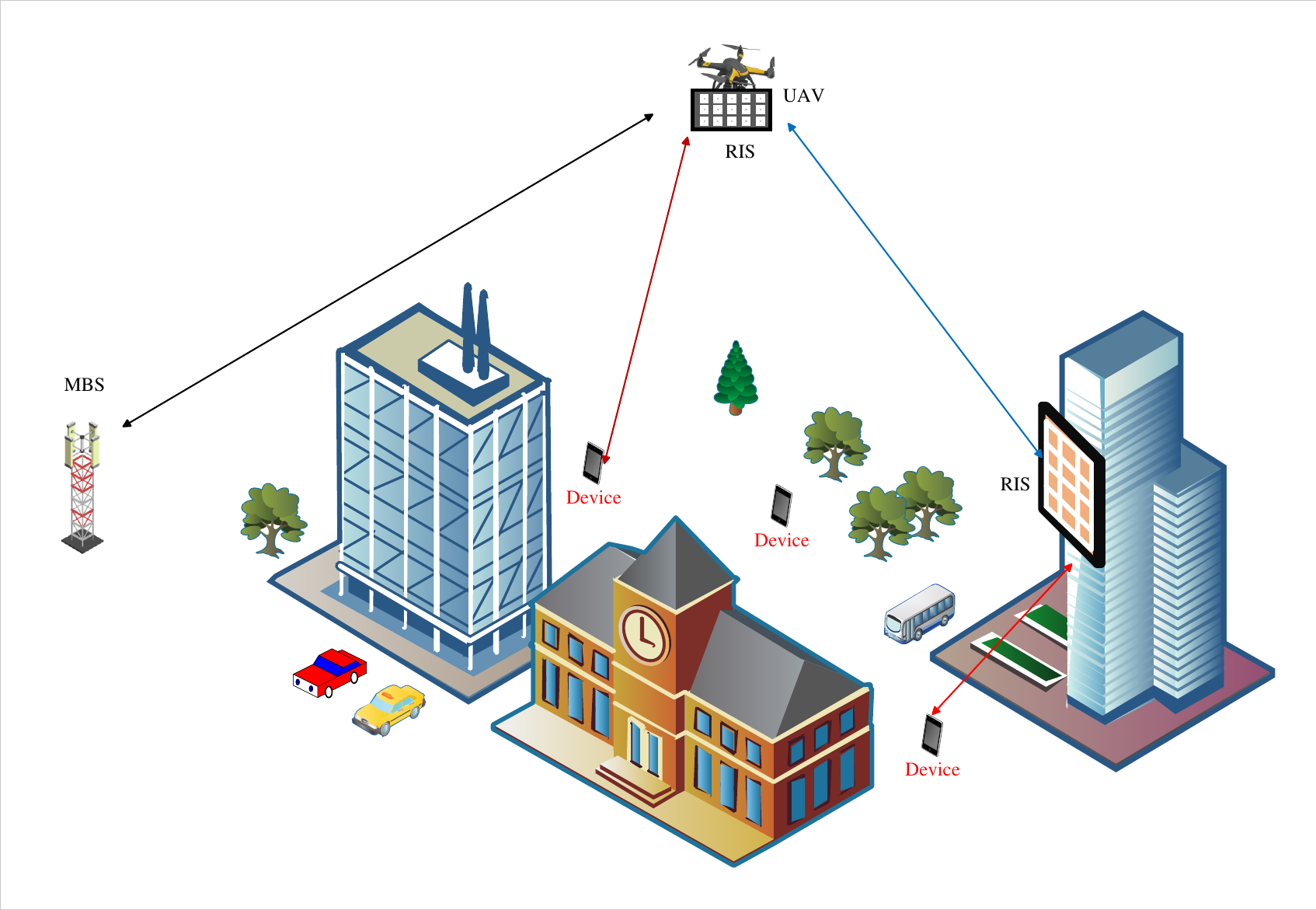}}
\caption{RIS-assisted and UAV-aided Network.}
\label{sys}
\end{figure*}

Motivated by these limitations, we propose an adaptive soft actor-critic (ASAC) framework that leverages UAV-mounted RIS (flying RIS) alongside ground-based RIS to boost network performance. Specifically, the ASAC model utilizes an adaptive sparse transformer with attentive feature refinement to address the complexities of joint optimization in real-time, focusing on the following motivations:
\begin{itemize}
    \item Enhanced Network Performance: Integrating ground and flying RIS can substantially improve connectivity, yet this architecture remains underexplored in existing research.
    \item Mitigating UAV Jitter: UAV movements introduce signal quality fluctuations that degrade communication. Our ASAC approach considers these impacts and dynamically adjusts system parameters without relying on outdated CSI.
    \item User Fairness: To achieve fair access across users in multi-user UAV-RIS networks, our approach maximizes the minimum user rate, ensuring balanced performance even in dense networks.
\end{itemize}
Based on the above motivations, we summarize our contributions as follows:
\begin{itemize}
  \item We propose a UAV-assisted RIS communication, where we have a ground RIS and flying RIS. Multiple IoT nodes are served using the proposed network. We aim at improving the fairness among nodes under the impact of UAV jitter, we formulate the optimization problem as maximization of the minimum rate of the node which is nonconvex.
  \item The problem is divided into three subproblems, beamforming subproblem, phase shift subproblem, and UAV trajectory subproblem. Then we propose an adaptive soft actor-critic (ASAC) framework in which we build adaptive sparse transformers with attentive feature refinement (ASTAFER) to design the agents of the proposed ASAC.
  \item Simulation results have shown that the proposed ASAC has outperforms the conventional soft actor-critic (SAC).
\end{itemize}
The remainder of the paper is organized as follows: In Section II, we introduce the system model and problem formulation. Section III includes the details of the proposed ASAC framework. In Section IV, we present the simulation results. Finally, in Section V, we have the conclusions.

\section{System Model and Problem Formulation}

We consider a UAV-aided, RIS-assisted IoT network as illustrated in Fig. \ref{sys}. In this setup, a UAV equipped with a reconfigurable intelligent surface (RIS) supports communication to remote IoT nodes, which are otherwise inaccessible from the base station (BS), denoted as $B$. To aid these blocked ground nodes, a ground-based RIS (denoted $R$) reflects the BS's transmitted signals towards the UAV-mounted RIS, enhancing the communication link.

The BS is positioned at coordinates $\bm{c}_B = [y_B, z_B]^T$ with a height $H_B$. We assume all communication links are secure and operate with stringent latency requirements \cite{bb16}. The set of IoT nodes is denoted as $\mathcal{K} = \{1, 2, \dots, K\}$, with the coordinates of each node $k \in \mathcal{K}$ represented by $\bm{c}_k = [x_k, y_k]^T$; each node is equipped with a single antenna.

The UAV's flight altitude is constant and represented by $H_U$. The UAV-mounted RIS, located at the horizontal coordinates $\bm{c}_U = [y_U, z_U]^T$ and altitude $H_U$, consists of $F = F_y \times F_z$ reflecting elements arranged in a uniform rectangular array (URA). The ground RIS, situated in the $x$-$z$ plane with horizontal coordinates $\bm{c}_R = [x_R, z_R]^T$ and altitude $R_z$, has $N = N_x \times N_z$ reflecting elements, also forming a URA.

Let the duration of each time slot be $\delta_t$; the finite UAV flight period $T$ is divided into $L$ time slots, where $T = L \delta_t$ and $l \in \mathcal{L} = \{1, \dots, L\}$. Thus, the coordinates of the UAV-mounted RIS, ground RIS, and IoT nodes at time slot $l$ are denoted as $\bm{c}_U[l] = [x_U[l], y_U[l]]^T$, $\bm{c}_R[l] = [x_R, z_R]^T$, and $\bm{c}_k[l] = [x_k, y_U]^T$, respectively. The UAV's trajectory follows the time-varying horizontal coordinates $\bm{c}_U[l]$.

The following constraints govern the UAV's movement \cite{bb22}:
\begin{equation} \label{eq1}
\| \bm{c}_U[l+1] - \bm{c}_U[l] \|^2 \le D^2, \quad l = 1, \dots, L - 1,
\end{equation}
\begin{equation} \label{eq2}
\| \bm{c}_U[L] - \bm{c}_U[0] \|^2 \le D^2,
\end{equation}
\begin{equation} \label{eq3}
\bm{c}_U[0] \equiv [0, 0, H_U]^T,
\end{equation}
where $D = \delta_t v_{\max}$ represents the maximum horizontal distance the UAV can travel within each time slot, and $v_{\max}$ is the UAV's maximum speed. Here, $\bm{c}_U[0]$ and $\bm{c}_U[L]$ are the predetermined initial and final locations of the UAV.

The passive beamforming matrices of the UAV-mounted and ground RIS at time slot $l$ are denoted as:
\begin{equation}
\Theta_U[l] = \mathrm{diag}\left( e^{j \theta_{U,1}[l]}, e^{j \theta_{U,2}[l]}, \dots, e^{j \theta_{U,F}[l]} \right),
\end{equation}
\begin{equation}
\Theta_R[l] = \mathrm{diag}\left( e^{j \theta_{R,1}[l]}, e^{j \theta_{R,2}[l]}, \dots, e^{j \theta_{R,N}[l]} \right),
\end{equation}
where $\theta_{U,f}[l]$ and $\theta_{R,m}[l]$ are the phase shifts for the reflection elements of the UAV-mounted and ground RIS, respectively, with $f \in \{1, \dots, F\}$ and $m \in \{1, \dots, N\}$. These phase shifts lie within the range $[0, 2\pi]$.

\subsection{Channel Model}

Let $\bm{h}_{kR} \in \mathbb{C}^{N \times 1}$, $\bm{h}_{kU} \in \mathbb{C}^{F \times 1}$, $\bm{H}_{RU} \in \mathbb{C}^{N \times F}$, and $\bm{h}_{UB} \in \mathbb{C}^{F \times 1}$ represent the channel gains between the node $k$ and the ground RIS, the node $k$ and the flying RIS, the ground RIS and the flying RIS, and the flying RIS and the BS, respectively. The reverse-order channel coefficients, representing the links from the BS to the ground node, are denoted as $\bm{h}_{BU} \in \mathbb{C}^{1 \times F}$, $\bm{H}_{UR} \in \mathbb{C}^{F \times N}$, $\bm{h}_{Rk} \in \mathbb{C}^{1 \times N}$, and $\bm{h}_{Uk} \in \mathbb{C}^{1 \times F}$. We assume that all communication links follow a Rician fading channel model.

In the downlink, the channels $\bm{h}_{BU} \in \mathbb{C}^{1 \times F}$, $\bm{H}_{UR} \in \mathbb{C}^{F \times N}$, $\bm{h}_{Rk} \in \mathbb{C}^{1 \times N}$, and $\bm{h}_{Uk} \in \mathbb{C}^{1 \times F}$ are given by:
\begin{equation}\label{eq4}
\bm{h}_{BU}[l] = \sqrt{\frac{\zeta_{BU}}{1 + \zeta_{BU}}} \bm{h}_{BU}^{\mathrm{LoS}}[l] + \sqrt{\frac{1}{1 + \zeta_{BU}}} \bm{h}_{BU}^{\mathrm{NLoS}},
\end{equation}
\begin{equation}\label{eq5}
\bm{H}_{UR}[l] = \sqrt{\frac{\zeta_{UR}}{1 + \zeta_{UR}}} \bm{H}_{UR}^{\mathrm{LoS}}[l] + \sqrt{\frac{1}{1 + \zeta_{UR}}} \bm{H}_{UR}^{\mathrm{NLoS}},
\end{equation}
\begin{equation}\label{eq6}
\bm{h}_{Rk}[l] = \sqrt{\frac{\zeta_{Rk}}{1 + \zeta_{Rk}}} \bm{h}_{Rk}^{\mathrm{LoS}}[l] + \sqrt{\frac{1}{1 + \zeta_{Rk}}} \bm{h}_{Rk}^{\mathrm{NLoS}},
\end{equation}
\begin{equation}\label{eq7}
\bm{h}_{Uk}[l] = \sqrt{\frac{\zeta_{Uk}}{1 + \zeta_{Uk}}} \bm{h}_{Uk}^{\mathrm{LoS}}[l] + \sqrt{\frac{1}{1 + \zeta_{Uk}}} \bm{h}_{Uk}^{\mathrm{NLoS}},
\end{equation}
where $\zeta_i$, $i \in \{BU, UR, Rk, Uk\}$ represents the Rician factor for each communication link. The terms $\bm{h}_i^{\mathrm{LoS}}[l]$ and $\bm{h}_i^{\mathrm{NLoS}}$ denote the line-of-sight (LoS) and non-line-of-sight (NLoS) components, respectively, with $\bm{h}_i^{\mathrm{NLoS}}$ being independently and identically distributed (i.i.d.) following the circularly symmetric complex Gaussian distribution $\mathcal{CN}(0,1)$.

The LoS component $\bm{H}_{UR}^{\mathrm{LoS}}[l]$ is given by:
\begin{equation}\label{eq8}
\bm{H}_{UR}^{\mathrm{LoS}}[l] = \bm{a}_{x}[l] \bm{a}_{z}^T[l],
\end{equation}
where $\bm{a}_{x}[l]$ and $\bm{a}_{z}[l]$ are the steering vectors defined as:
\begin{equation}\label{eq9}
\bm{a}_{x}[l] = \begin{bmatrix}
1 \\
e^{-j \frac{2\pi}{\lambda} \vartheta \cos \varphi_{UR}[l] \sin \phi_{UR}[l]} \\
\vdots \\
e^{-j \frac{2\pi}{\lambda} \vartheta (F_x - 1) \cos \varphi_{UR}[l] \sin \phi_{UR}[l]}
\end{bmatrix},
\end{equation}
\begin{equation}\label{eq10}
\bm{a}_{z}[l] = \begin{bmatrix}
1 \\
e^{-j \frac{2\pi}{\lambda} \vartheta \sin \varphi_{UR}[l] \sin \phi_{UR}[l]} \\
\vdots \\
e^{-j \frac{2\pi}{\lambda} \vartheta (F_z - 1) \sin \varphi_{UR}[l] \sin \phi_{UR}[l]}
\end{bmatrix}.
\end{equation}

Here, $\varphi_{UR}[l]$ and $\phi_{UR}[l]$ are the azimuth and elevation angles, $\vartheta$ is the antenna separation, and $\lambda$ is the wavelength. We have:
\begin{equation}
    \cos \varphi_{UR}[l] \sin \phi_{UR}[l] = \frac{x_R - x_U[l]}{d_{UR}[l]}, \\
\end{equation}
\begin{equation}
    \sin \varphi_{UR}[l] \sin \phi_{UR}[l] = \frac{R_z - H_U}{d_{UR}[l]},
\end{equation}
where the distance $d_{UR}[l]$ is given by:
\begin{equation}
d_{UR}[l] = \sqrt{\left\| \bm{c}_R - \bm{c}_U[l] \right\|^2 + \left(R_z - H_U\right)^2}.
\end{equation}

Since the UAV's stability cannot be fully guaranteed, we consider movement-induced jitter. This jitter results from mechanical interactions between various UAV components, as well as from pitch, rotation, yaw movements, and external factors like wind. Such jitter can lead to imperfect channel state information (CSI) estimation and cause instability in communication links \cite{bb11}. Following the approach in \cite{bb22}, jitter in the UAV's pitch and roll is modeled through variations in the elevation angle, while jitter in the yaw is captured by variations in the azimuth angle.
In this work, we define the following set of angles to represent the UAV and RIS orientation at each time slot $l$:
\begin{equation}\label{eq15}
\begin{aligned}
\varpi_{BU}[l] &= \tilde{\varpi}_{BU}[l] + \Delta \varpi_{BU}[l], \\
\phi_{BU}[l] &= \tilde{\phi}_{BU}[l] + \Delta \phi_{BU}[l], \\
\Xi_{BU}[l] &= \left\{
\begin{array}{l}
\Delta \varpi_{BU}[l], \Delta \phi_{BU}[l] \in \mathbb{R} \mid \\
\left( \Delta \varpi_{BU}[l] \right)^2 + \left( \Delta \phi_{BU}[l] \right)^2 \le \psi_{BU}^2
\end{array}
\right\},
\end{aligned}
\end{equation}
\begin{equation}\label{eq16}
\begin{aligned}
\varpi_{UR}[l] &= \tilde{\varpi}_{UR}[l] + \Delta \varpi_{UR}[l], \\
\phi_{UR}[l] &= \tilde{\phi}_{UR}[l] + \Delta \phi_{UR}[l], \\
\Xi_{UR}[l] &= \left\{
\begin{array}{l}
\Delta \varpi_{UR}[l], \Delta \phi_{UR}[l] \in \mathbb{R} \mid \\
\left( \Delta \varpi_{UR}[l] \right)^2 + \left( \Delta \phi_{UR}[l] \right)^2 \le \psi_{UR}^2
\end{array}
\right\},
\end{aligned}
\end{equation}
\begin{equation}\label{eq17}
\begin{aligned}
\varpi_{Uk}[l] &= \tilde{\varpi}_{Uk}[l] + \Delta \varpi_{Uk}[l], \\
\phi_{Uk}[l] &= \tilde{\phi}_{Uk}[l] + \Delta \phi_{Uk}[l], \\
\Xi_{Uk}[l] &= \left\{
\begin{array}{l}
\Delta \varpi_{Uk}[l], \Delta \phi_{Uk}[l] \in \mathbb{R} \mid \\
\left( \Delta \varpi_{Uk}[l] \right)^2 + \left( \Delta \phi_{Uk}[l] \right)^2 \le \psi_{Uk}^2
\end{array}
\right\}.
\end{aligned}
\end{equation}
Here, $\tilde{\varpi}_{BU}[l]$ and $\Delta \varpi_{BU}[l]$ represent the estimated azimuth and its uncertainty for the link between the BS and the flying RIS, respectively, while $\tilde{\phi}_{BU}[l]$ and $\Delta \phi_{BU}[l]$ represent the estimated elevation and its uncertainty. Sets $\Xi_{BU}[l]$, $\Xi_{UR}[l]$, and $\Xi_{Uk}[l]$ capture all possible uncertainties in azimuth and elevation angles, bounded by upper limits $\psi_{BU}^2$, $\psi_{UR}^2$, and $\psi_{Uk}^2$ for the BS-to-flying RIS, flying RIS-to-ground RIS, and flying RIS-to-node communication links, respectively.

Similarly, the channels $\bm{h}_{BU} \in \mathbb{C}^{1 \times F}$, $\bm{h}_{Rk}^H \in \mathbb{C}^{1 \times N}$, and $\bm{h}_{Uk}^H \in \mathbb{C}^{1 \times F}$ are defined. For the indirect link between the BS and the node (via the flying RIS and ground RIS), the distance-based path loss is given by:
\begin{equation}\label{eq11}
\mathcal{D}_{URk}[l] = \sqrt{\beta \left( d_{BU}[l] d_{UR}[l] d_{Rk} \right)^{-\alpha}},
\end{equation}
where $\beta$ is the path loss at the reference distance $D_0 = 1$, and $\alpha$ is the path loss exponent over the entire indirect link between the UAV $U$ and node $k$ via the RIS $R$. The distance $d_{Rk}$ is given by:
\begin{equation}\label{eq12}
d_{Rk} = \sqrt{\left\| \bm{c}_R - \bm{c}_k \right\|^2 + R_z^2}.
\end{equation}

The distance-based path loss for the direct links between the ground RIS $R$ and the node $k$, and the flying RIS and the node $k$, are:
\begin{equation}\label{eq13}
\mathcal{D}_{Rk} = \sqrt{\beta d_{Rk}^{- \frac{\varepsilon}{2}}},
\end{equation}
\begin{equation}\label{eq14}
\mathcal{D}_{Uk} = \sqrt{\beta d_{Uk}^{- \frac{\varepsilon}{2}}},
\end{equation}
where $\varepsilon$ is the path loss exponent of the direct communication links between the ground RIS $R$ and the node $k$, and between the flying RIS and the node $k$.

\subsection{Signal Model}
In the downlink (DL) transmission, the signal is transmitted from the base station (BS) $B$ to the node $k$ via the flying RIS $U$. We denote the beamforming vector for node $k$ at the BS as $\bm{w}_k^d[l]$, subject to the following power constraint:
\begin{equation}\label{eq35}
\frac{1}{L} \sum_{l=1}^L \sum_{k \in \mathcal{K}} \|\bm{w}_k^d[l]\|^2 \le P^d,
\end{equation}
where $P^d$ is the maximum total transmit power of the BS.
The transmitted signal at $B$ is given by:
\begin{equation}\label{eq37}
\bm{x}_B[l] = \sum_{k \in \mathcal{K}} \bm{w}_k^d[l] s_k^d[l] + \bm{n},
\end{equation}
where $s_k^d[l]$ is the data symbol for node $k$ at time slot $l$ and $\bm{n}$ is the noise vector.
The signal-to-interference-plus-noise ratio (SINR) at the receiver node $k$ is expressed as in \eqref{eq38} on top of next page.

\begin{figure*}
    \begin{equation}\label{eq38}
\gamma_k^d[l] = \frac{\left| \left( \bm{H}_{BU}[l] \bm{\Theta}_U^d[l] \bm{H}_{UR}^H[l] \bm{\Theta}_R^d[l] \bm{h}_{Rk}^H \mathcal{D}_{BURk}[l] + \bm{H}_{BU}[l] \bm{\Theta}_U^d[l] \bm{h}_{Uk}^H[l] \mathcal{D}_{BUk}[l] \right) \bm{w}_k^d[l] \right|^2}{\sum\limits_{j \in \mathcal{K}, j \neq k} \left| \left( \bm{H}_{BU}[l] \bm{\Theta}_U^d[l] \bm{H}_{UR}^H[l] \bm{\Theta}_R^d[l] \bm{h}_{Rj}^H \mathcal{D}_{BURj}[l] + \bm{H}_{BU}[l] \bm{\Theta}_U^d[l] \bm{h}_{Uj}^H[l] \mathcal{D}_{BUj}[l] \right) \bm{w}_j^d[l] \right|^2 + \sigma^2},
\end{equation}
\end{figure*}
where $\sigma^2$ denotes the noise variance, $\bm{H}_{BU}[l]$ and $\bm{H}_{UR}[l]$ are channel matrices, $\bm{h}_{Rk}[l]$ and $\bm{h}_{Uk}[l]$ are channel vectors, and $\mathcal{D}_{BURk}[l]$ and $\mathcal{D}_{BUk}[l]$ represent the path loss components. The denominator represents interference from all other users $j \in \mathcal{K}, j \neq k$.
The achievable rate for node $k$ is then given by:
\begin{equation}\label{eq39}
r_k^d[l] = \log_2 \left( 1 + \gamma_k^d[l] \right).
\end{equation}
For the uplink (UL) case, let $\bm{w}_k^u[l]$ represent the beamforming vector for node $k$ at the RIS, with a power constraint defined as:
\begin{equation}\label{eq40}
\frac{1}{L} \sum_{l=1}^L \sum_{k \in \mathcal{K}} \|\bm{w}_k^u[l]\|^2 \le P^u,
\end{equation}
where $P^u$ is the maximum total transmit power for the UL. The SINR for node $k$ in the UL case as in \eqref{eq42} on top of next page.
\begin{figure*}
    \begin{equation}\label{eq42}
\gamma_k^u[l] = \frac{\left| \left( \bm{H}_{UB}[l] \bm{\Theta}_U^u[l] \bm{h}_{kU}^H[l] \mathcal{D}_{kUB}[l] + \bm{H}_{UB}[l] \bm{\Theta}_U^u[l] \bm{H}_{RU}^H[l] \bm{\Theta}_R^u[l] \bm{h}_{kR}^H \mathcal{D}_{kRUB}[l] \right) \bm{w}_k^u[l] \right|^2}{\sum\limits_{j \in \mathcal{K}, j \neq k} \left| \left( \bm{H}_{UB}[l] \bm{\Theta}_U^u[l] \bm{h}_{jU}^H[l] \mathcal{D}_{jUB}[l] + \bm{H}_{UB}[l] \bm{\Theta}_U^u[l] \bm{H}_{RU}^H[l] \bm{\Theta}_R^u[l] \bm{h}_{jR}^H \mathcal{D}_{jRUB}[l] \right) \bm{w}_j^u[l] \right|^2 + \sigma^2},
\end{equation}
\end{figure*}

where $\bm{H}_{UB}[l]$ and $\bm{H}_{RU}[l]$ are channel matrices, and $\bm{h}_{kU}[l]$ and $\bm{h}_{kR}[l]$ are channel vectors.
The achievable data rates for node $k$ in DL and UL are then expressed as:
\begin{equation}\label{eq43}
R_k^d = \frac{1}{L} \sum_{l=1}^L \log_2 r_k^d[l],
\end{equation}
\begin{equation}\label{eq44}
R_k^u = \frac{1}{L} \sum_{l=1}^L \log_2 r_k^u[l] ,
\end{equation}
where $r_k^d[l]$ and $r_k^u[l]$ represent the instantaneous rates at node $k$ in the DL and UL, respectively. The total rate is then defined as
\begin{equation}
    R = \frac{1}{L}\sum_{l=1}^L\left[\varepsilon r_k^d[l] + \left(1 - \varepsilon\right)r_k^u[l]\right]
\end{equation}

\subsection{Problem Formulation}
Our objective is to maximize the minimum rate of the node in the DL and UL by jointly optimizing the beamforming $\mathbf{W}$, the phase shift $\Phi$, and the UAV trajectory $\mathbf{Q}$. Thus, we define our optimization problem as follows
\begin{subequations}\label{eq50:main}
\begin{align}
& \mathop {\max }\limits_{{\mathbf{W}},{\mathbf{Q}},{\Phi _U},{\Phi _R}} \mathop {\min }\limits_{\forall k \in {\cal K}}&& R    &   & \tag{\ref{eq50:main}} \\
& \text{s.t.}&& 0 \le {\theta _{U,f}}\left[ l \right] \le 2\pi ,\label{eq50:b} \\
&             && 0 \le {\theta _{R,m}}\left[ l \right] \le 2\pi,\label{eq50:c}\\
&             &&\eqref{eq1},\eqref{eq2},\eqref{eq3},\eqref{eq35},\eqref{eq40}.\nonumber
\end{align}
\end{subequations}

\section{Proposed Adaptive Soft Actor-Critic}
The soft actor-critic (SAC) \cite{bb23,bb24} algorithm optimizes a stochastic policy using an off-policy approach, bridging the gap between stochastic policy optimization and methods similar to DDPG. SAC uses the clipped double-Q technique and, thanks to its inherently stochastic policy, gains an advantage similar to that provided by target policy smoothing.

Generally, to apply DRL, we reformulate the optimization problem as a Markov decision process (MDP) with specifically defined states, actions, and rewards:
\begin{itemize}
    \item \textbf{State:} The environment state at each timeslot $t$ is represented as $s_t = [\bm{q}_U^t, \mathbf{H}^t]$, where $\bm{q}_U^t$ is the UAV's position at timeslot $t$, and $\mathbf{H}^t$ denotes the channel state information (CSI) between IoT nodes and different terminals.

    \item \textbf{Action:} The UAV agent's action at timeslot $t$ is defined as $a_t = [v_t, \theta_t, \{k \in \mathcal{K} \mid s_k^t = 1\}, \mathbf{W}, \boldsymbol{\Phi}_U, \boldsymbol{\Phi}_R]$. Here, $v_t$ and $\theta_t$ represent the UAV's average speed and horizontal direction, respectively; $\{k \in \mathcal{K} \mid s_k^t = 1\}$ specifies the transmission scheduling for IoT devices; $\mathbf{W}$ is the beamforming vector, while $\boldsymbol{\Phi}_U$ and $\boldsymbol{\Phi}_R$ are the phase shift matrices of the UAV-mounted and ground RIS elements, respectively. The parameters $v_t$, $\theta_t$, $s_k^t$, $\mathbf{W}$, $\boldsymbol{\Phi}_U$, and $\boldsymbol{\Phi}_R$ must satisfy constraints (10b), (10c), (10f)-(10h).

    \item \textbf{Reward:} To maximize the minimum achievable rate $R$ for user $k \in \mathcal{K}$, the reward function at timeslot $t$ is designed as:
    \begin{equation}
      r_t = \min_{k \in \mathcal{K}} R_k^t + \Delta_{\text{penalty}},
    \end{equation}
    where $\Delta_{\text{penalty}}$ applies if the UAV moves outside the designated region, violating constraints.
\end{itemize}

\subsection{Proposed ASAC Algorithm}
\begin{figure}
    \centering
    \includegraphics[width=3.2in]{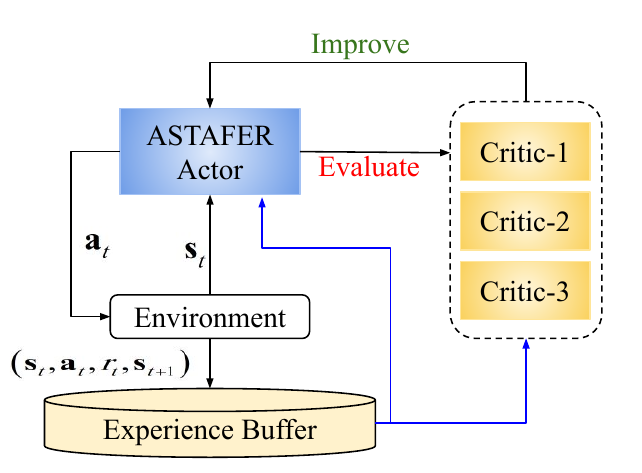}
    \caption{Structure of the proposed ASAC.}
    \label{fig:asac}
\end{figure}
We present the proposed ASAC framework, which combines an SAC algorithm and transformer model for controlling  beamforming, RIS phase shifts, and UAV trajectory. This algorithm framework, illustrated in Fig. \ref{fig:asac}, includes the following key features in the SAC component
\begin{enumerate}
    \item Entropy Maximization: This element ensures robust exploration, helping the algorithm to sample diverse actions.
    \item Adaptive Sparse Transformer Actor with Attentive Feature Refinement (ASTAFER): The actor network uses an ASTAFER architecture for efficient action selection, balancing model sparsity with attention-based refinement to prioritize relevant features.
    \item Multiple Critic Networks: The critic section consists of three deep neural networks (DNNs), enabling a more robust assessment of action values for improved stability.
    \item Prioritized Experience Replay (PER): PER is used to enhance convergence speed and robustness during training.
\end{enumerate}
\textbf{Soft Value Functions:} Unlike conventional DRL methods such as DQN and DDPG that learn deterministic policies, SAC optimizes a stochastic policy aiming to maximize both cumulative reward and expected entropy. The objective function is
\begin{equation}
J(\pi) = \sum_{t=0}^T \mathbb{E}_{(s_t, a_t) \sim \rho^\pi} \left[ r(s_t, a_t) + \zeta H(\pi(\cdot | s_t)) \right],
\end{equation}
where $\rho^\pi$ is the state-action distribution under policy $\pi$, $r(s_t, a_t)$ represents the reward, and $\zeta$ is a temperature parameter balancing entropy with the reward. The entropy term $H(\pi(\cdot | s_t)) = -\log \pi(a_t | s_t)$ encourages exploration. Thus, the entropy-augmented objective function $J(\pi)$ is:
\begin{equation}
J(\pi) = \sum_{t=0}^T \mathbb{E} \left[ r(s_t, a_t) - \zeta \log \pi(a_t | s_t) \Big| \pi \right].
\end{equation}
In the policy evaluation step, given an initial state $s$ and action $a$, the soft Q-value function is derived as
\begin{equation}
Q(s, a) = \mathbb{E} \left[ \sum_{t=1}^T \gamma^t \left[ r_t - \zeta \log \pi(a_t | s_t) \right] \Big| s_0 = s, a_0 = a, \pi \right],
\end{equation}
where $r_t$ is the reward, which depends on the UAV's trajectory, beamforming, and phase shifts. The soft value function is further derived from the Bellman backup as follows
\begin{equation}
V(s_t) = \mathbb{E}_{a_t \sim \pi} \left[ Q(s_t, a_t) - \zeta \log \pi(a_t | s_t) \right].
\end{equation}
Since iterating the value function until convergence is computationally expensive, we use deep neural networks (DNNs) to approximate the soft Q-function in the critic networks and the policy in the ASTAFER-based actor network. We apply stochastic gradient descent to alternately update the parameters in both the actor and critic networks.\\
\textbf{Critic Part:} In the critic part, three Q-networks ($\psi_1$, $\psi_2$, and $\psi_3$) are employed to estimate the state-action values from different perspectives, ensuring stable performance under complex scenarios. Additionally, a value network $\phi$ estimates state values, and a target value network $\bar{\phi}$ maintains an exponential moving average of $\phi$. The soft value function is trained by minimizing the squared residual error
\begin{equation}\label{eq38}
\begin{aligned}
J_V(\phi) = \mathbb{E}_{s_t \sim B} \Bigg[ \frac{1}{2} \Big( V_\phi(s_t)
&- \mathbb{E}_{a_t \sim \pi_\theta} \big[ Q_\psi(s_t, a_t) \\
&- \zeta \log \pi_\theta(a_t | s_t) \big] \Big)^2 \Bigg],
\end{aligned}
\end{equation}
where $B$ is a replay buffer that stores experience tuples $(s_t, a_t, r_t, s_{t+1})$. The gradient of \eqref{eq38} can be estimated as:
\begin{equation}
\begin{aligned}
\hat{\nabla}_\phi J_V(\phi) = \nabla_\phi V_\phi(s_t) \times
\Big( V_\phi(s_t) &- Q_\psi(s_t, a_t) \\
&+ \zeta \log \pi_\theta(a_t | s_t) \Big),
\end{aligned}
\end{equation}
The parameters of the value network $\phi$ are then updated as:
\begin{equation}\label{eq40}
\phi \leftarrow \phi - \lambda_V \hat{\nabla}_\phi J_V(\phi),
\end{equation}
where $\lambda_V \geq 0$ is the learning rate for the value network. The target value network $\bar{\phi}$ is updated by:
\begin{equation}\label{eq41}
\bar{\phi} \leftarrow \tau \phi + (1 - \tau) \bar{\phi},
\end{equation}
with $\tau \in [0, 1]$ as the target network update rate. The soft Q-functions are optimized by minimizing the soft Bellman residual
\begin{equation}
J_Q(\psi) = \mathbb{E}_{(s_t, a_t) \sim B} \left[ \frac{1}{2} \left( Q_\psi(s_t, a_t) - \hat{Q}(s_t, a_t) \right)^2 \right],
\end{equation}
where
\begin{equation}
\hat{Q}(s_t, a_t) = r(s_t, a_t) + \gamma \mathbb{E}_{s_{t+1} \sim p} \left[ V_{\bar{\phi}}(s_{t+1}) \right].
\end{equation}
The gradient is given by
\begin{equation}
\begin{aligned}
\hat{\nabla}_\psi J_Q(\psi) = \nabla_\psi Q_\psi(s_t, a_t) \times
\Big( Q_\psi(s_t, a_t) &- r(s_t, a_t) \\
&- \gamma V_{\bar{\phi}}(s_{t+1}) \Big).
\end{aligned}
\end{equation}
The Q-network parameters $\psi_i$ for $i \in \{1, 2, 3\}$ are updated by
\begin{equation}\label{eq45}
\psi_i \leftarrow \psi_i - \lambda_Q \hat{\nabla}_{\psi_i} J_Q(\psi_i),
\end{equation}
where $\lambda_Q \geq 0$ is the learning rate for the Q-networks.\\
\textbf{Actor Part:} The actor network, implemented as an ASTAFER, generates the mean and standard deviation for the action distribution, selecting actions based on beamforming, phase shift tuning, and UAV trajectory adjustments. The policy is updated by minimizing the expected Kullback-Leibler (KL) divergence, resulting in an improved policy:
\begin{equation}
\pi_{\text{new}} = \arg \min_{\pi_\theta \in \Pi} D_{\text{KL}} \left( \pi_\theta(\cdot | s_t) \parallel \frac{\exp(Q_{\pi_{\text{old}}}(s_t, \cdot) / \zeta)}{Z_{\pi_{\text{old}}}(s_t)} \right),
\end{equation}
where $\Pi$ is the feasible policy set, $D_{\text{KL}}(\cdot)$ is the KL divergence, and $Z_{\pi_{\text{old}}}(s_t)$ normalizes the distribution. A DNN parameterized by $\theta$ approximates the policy function $\pi_\theta(\cdot | s_t)$ by minimizing:
\begin{equation}
J_\pi(\theta) = \mathbb{E}_{s_t \sim B} \left[ \mathbb{E}_{a_t \sim \pi_\theta} \left[ \zeta \log (\pi_\theta(a_t | s_t)) - Q_\psi(s_t, a_t) \right] \right].
\end{equation}
The policy is reparameterized using:
\begin{equation}
a_t = f_\theta(\epsilon_t; s_t),
\end{equation}
where $\epsilon_t$ is input noise. The output of $f_\theta$ includes mean $f_\mu^\theta$ and standard deviation $f_\sigma^\theta$, so we rewrite as:
\begin{equation}
a_t = f_\mu^\theta\left(s_t\right) + \epsilon_t \odot f_\sigma^\theta\left(s_t\right),
\end{equation}
with $\odot$ as the element-wise product. Finally, the objective is reformulated as
\begin{equation}
J_\pi(\theta) = \mathbb{E}_{s_t \sim B, \epsilon_t \sim \mathcal{N}} \left[\begin{array}{l}
\log {\pi _\theta }({f_\theta }({\epsilon_t};{s_t})|{s_t})\\
 - {Q_\psi }({s_t},{f_\theta }({\epsilon_t};{s_t}))
\end{array} \right],
\end{equation}
with gradients approximated for updating the policy network parameters.

\textbf{RPER Technique:} SAC, as an off-policy algorithm, utilizes experience replay to enhance learning efficiency. Traditionally, during parameter updates, mini-batches of data are drawn uniformly and randomly from the replay buffer. However, as the volume of stored transitions increases, purely random sampling can lead to instability in training or even hinder convergence. To address this, the RPER mechanism combines aspects of emphasizing recent experience (ERE) \cite{bb25} and prioritized experience replay (PER) \cite{bb26}, specifically designed to stabilize the SAC training process. The RPER technique operates in two primary ways:

First, the sampling range is incrementally restricted to prioritize more recent data points. Let $\mathbf{B}$ represent the number of mini-batch updates scheduled for the current phase. For the $u$-th update (where $1 \leq \beta \leq \mathbf{B}$), the most recent $c_u$ data points are sampled uniformly accoding to the following
\begin{equation}
c_\beta = \max \left\{ \mathcal{B}_{\text{max}} \cdot \eta^{\frac{\beta}{1000U}}, c_{\text{min}} \right\},
\end{equation}
where $\mathcal{B}_{\text{max}}$ is the maximum capacity of the replay buffer, $\eta \in (0,1]$ is a hyperparameter that controls the emphasis on recent experiences, and $c_{\text{min}}$ is the minimum allowable size of $c_\beta$. This strategy provides a more accurate approximation of value functions around recently-visited states while still accounting for states encountered further in the past.

Second, within this defined sampling range, priority sampling is applied, where the probability of selecting a data point is based on the absolute temporal-difference (TD) error. As TD error directly impacts the critic network's updates, selecting points with higher TD error can yield more informative updates. Let $D_{c_\beta}$ represent the $c_\beta$ most recent data points in the buffer. The probability of sampling a data point $i$ is calculated as
\begin{equation}
P(i) = \frac{p_i^\alpha}{\sum_j p_j^\alpha}, \quad i, j \in D_{c_\beta},
\end{equation}
where $p_i$ is the priority value of the $i$-th data point, determined by its absolute TD error.

\begin{algorithm}
\caption{Proposed Solution: ASAC for Joint Beamforming, Phase Shift, and UAV Trajectory}
\begin{algorithmic}[1]
\STATE Initialize parameters $\theta$, $\psi$, $\phi$, $\bar{\phi}$
\STATE Initialize experience replay buffer $\mathcal{B}$
\FOR{each episode}
    \STATE Initialize environment and observe initial state $s_0$
    \FOR{timeslot $t = 1, 2, \dots, T$}
        \STATE Observe state $s_t$ and take action $a_t \sim \pi_\theta(\cdot | s_t)$
        \STATE Jointly optimize beamforming, phase shifts of the RIS, and UAV trajectory based on the chosen action $a_t$
        \STATE Obtain a new state $s_{t+1}$ and a reward $r_t$
        \IF{the UAV flies outside the boundary}
            \STATE $r_t = r_t + \Delta_{\text{penalty}}$, where $\Delta_{\text{penalty}}$ is a given penalty, and reset the UAV's movement, updating $s_{t+1}$ and $r_t$
        \ENDIF
        \STATE Store experience $(s_t, a_t, r_t, s_{t+1})$ in the replay buffer $B$
        \IF{$t = T$ (terminal state)}
            \STATE Sample a batch of experiences using the RPER technique;
            \STATE Update the parameters of the value network as defined in \eqref{eq40};
            \STATE Update the parameters of the two soft Q-networks as defined in \eqref{eq45};
            \STATE Update the parameters of the policy network as
            \begin{equation}\label{eq40}
                   \theta \leftarrow \theta - \pi_V \hat{\nabla}_\phi J_V(\theta),
            \end{equation}
            \STATE Update the parameters of the target value network as defined in \eqref{eq41}
        \ENDIF
    \ENDFOR
\ENDFOR
\end{algorithmic}
\end{algorithm}

\subsection{Actor and Critics Architecture}
The proposed actor network is based on the ASTAFER model in \cite{bb27}. The details of the ASTAFER are given as follows:
\subsubsection{ASTAFER-based actor network}
In designing an ASTAFER-based actor network within the soft actor-critic (SAC) framework for optimizing beamforming $\mathbf{W}$, phase shifts $\boldsymbol{\Phi}_U$ and $\boldsymbol{\Phi}_R$, and UAV trajectory $\mathbf{Q}$, the Adaptive Sparse Transformer with Attentive Feature Refinement (ASTAFER) model is structured to handle high-dimensional and structured action spaces. This architecture allows for dynamic focus on relevant action components, thus enhancing stability and efficiency in decision-making. The following describes the architecture of the ASTAFER-based actor network.

The input state, which encapsulates information about the environment (e.g., current positions, communication channel states, and initial beamforming and phase shift parameters), is processed to produce embeddings for beamforming vectors $\mathbf{W}$, phase shifts $\boldsymbol{\Phi}_U$ and $\boldsymbol{\Phi}_R$ of the RIS, and UAV trajectory $\mathbf{Q}$. These embeddings serve as the basis for constructing the query, key, and value matrices $Q$, $K$, and $V$, allowing the model to selectively attend to relevant features within the action space via the ASTAFER structure.
\begin{itemize}
\item Adaptive Sparse self-attention (ASSA) mechanism:
The ASSA mechanism in ASTAFER balances sparse and dense self-attention to manage the high-dimensional and structured action space, focusing on critical actions for effective control of $\mathbf{W}$, $\boldsymbol{\Phi}_U$, $\boldsymbol{\Phi}_R$, and $\mathbf{Q}$. The sparse self-attention (SSA) branch emphasizes high-relevance action parameters while filtering out less impactful actions through the following
\begin{equation}
\text{SSA} = \text{ReLU}^2\left(\frac{QK^T}{\sqrt{d}} + B\right),
\end{equation}
where SSA filters out low-relevance features, reducing noise and computational load by emphasizing only critical interactions within the network.
\item Dense self-attention (DSA): branch, in contrast, aggregates information from all action dimensions to ensure that key features across $\mathbf{W}$, $\boldsymbol{\Phi}_U$, $\boldsymbol{\Phi}_R$, and $\mathbf{Q}$ are comprehensively represented:
\begin{equation}
\text{DSA} = \text{Softmax}\left(\frac{QK^T}{\sqrt{d}} + B\right).
\end{equation}
The fusion of SSA and DSA enables the model to adaptively balance between focused exploration of critical actions and maintaining an overall stable policy, dynamically adjusting to changes in the environment.
\item Feature refinement feed-forward network (FRFN):
The FRFN module refines the feature representations, reducing redundancies and enhancing important characteristics along key action dimensions. Specifically, FRFN operates through the following transformation:
\begin{equation}
X'_{\text{out}} = \text{GELU}\left(W_2 \left( X'_1 \otimes F\left(\text{DWConv}(R(X'_2))\right) \right)\right),
\end{equation}
where $X'_1$ and $X'_2$ represent intermediate action features for $\mathbf{W}$, $\boldsymbol{\Phi}_U$, $\boldsymbol{\Phi}_R$, and $\mathbf{Q}$. The projection matrices $W_1$ and $W_2$ control the transformation process, while partial convolution ($PConv(\cdot)$) emphasizes essential features. Depthwise convolution ($DWConv(\cdot)$) captures dependencies along channel dimensions, refining the network's representation of actions in the actor space.
\item Policy Output:
The ASTAFER-based actor network generates a stochastic policy over the action space, outputting mean and standard deviation values for $\mathbf{W}$, $\boldsymbol{\Phi}_U$, $\boldsymbol{\Phi}_R$, and $\mathbf{Q}$:
\begin{equation}
\pi(a_t | s_t) \sim \mathcal{N}(\mu_{\theta}(s_t), \sigma_{\theta}(s_t)),
\end{equation}
where the adaptive attention mechanisms of ASSA and feature refinement by FRFN enable the model to dynamically focus on the most influential actions according to real-time network conditions.
\end{itemize}

\subsubsection{DNN-based Critic Network}
In the proposed ASAC, each critic (denoted as $\psi_1$, $\psi_2$, and $\psi_3$) is implemented using a DNN tailored to evaluate the actions generated by the ASTAFER-based actor network across the three primary dimensions: beamforming $\mathbf{W}$, phase shifts $\boldsymbol{\Phi}_U$ and $\boldsymbol{\Phi}_R$, and UAV trajectory $\mathbf{Q}$. The ASTAFER-based actor thus interacts with these critics to receive feedback on its action choices, enhancing the learning stability and robustness. Here's a detailed description of how these DNN-based critics can be structured and how they interact with the ASTAFER-based actor network.

Each critic network takes as input the current state, $s_t$, which represents environmental context (such as positions, channel states, and current network conditions), and the action output from the ASTAFER-based actor network, which is a combined vector of $\mathbf{W}$, $\boldsymbol{\Phi}_U$, $\boldsymbol{\Phi}_R$, and $\mathbf{Q}$. This action encapsulates the optimized parameters for beamforming, RIS phase shifts, and UAV trajectory that the actor network suggests. Each critic network then approximates the expected reward $Q(s_t, a_t)$ based on this input.

The architecture of each critic network is implemented as a separate DNN with five layers. While these networks share similar architectures, each critic may be slightly tuned to emphasize different aspects of the action space. Typically, each DNN critic begins with an input layer that concatenates the state $s_t$ and action $a_t$ vectors, including all action components: $\mathbf{W}$, $\boldsymbol{\Phi}_U$, $\boldsymbol{\Phi}_R$, and $\mathbf{Q}$. Following this, each DNN contains three fully connected hidden layers with nonlinear activations such as rectified linear unit (ReLU) and Gaussian error linear unit (GELU) to capture the complex interactions between the state and action spaces. Layer normalization is applied after each hidden layer to improve training stability. Finally, the output layer of each critic produces a single scalar $Q_{\psi_i}(s_t, a_t)$, representing the estimated value for the given state-action pair.

\section{Numerical Results}
In this section, we present our simulation results to evaluate the performance of the proposed framework. The simulation parameters follow the values shown in TABLE \ref{table2} unless otherwise stated.

The study is conducted within a 800 m $\times$ 800 m urban environment characterized by dense, high-rise buildings. A single UAV is tasked with collecting data from \( K \) IoT nodes, with an RIS mounted on a designated building located at \( \mathbf{c}_R = [360, 200, 80] \). The UAV begins its mission at \( \mathbf{c}_U(0) = [80, 80, 100] \) and maintains a fixed altitude of \( H_U = 100 \) m, flying at a maximum speed of \( v_{\max} = 20 \) m/s. The scenario is divided into 250 timeslots, each lasting \( \tau = 1 \) second. Parameters include a discount factor \( \gamma = 0.99 \), a learning rate of \( 1 \times 10^{-4} \), and a batch size of 128.

The simulations and training are run in x64 workstation with Intel(R) Core(TM) i9-10900K CPU@3.70GHz, RAM 64 GB, NVIDIA Quadro RTX 5000 GPU, and Windows 10 operating system. The version of Python is 3.12.3 and PyTorch version is 2.3.0.
\begin{table}
\centering
\caption{Simulation Parameters}
\label{table2}
\begin{tabular}{|l|l|}
  \hline
  \textbf{Parameter} & \textbf{Value} \\
  \hline
UAV altitude &	$H_U = 100$ m\\
\hline
$v_{max}$ & 20 m/s\\
\hline
Number of flying RIS elements & $F = 36$\\	
\hline
Flying RIS elements spacing & ${\vartheta _y} = {\vartheta _z} = \frac{\lambda }{4}$\\	
\hline
RIS altitude &	30 m\\
\hline
Number of RIS elements & $N = 64$\\	
\hline
RIS elements spacing & ${\vartheta _x} = {\vartheta _z} = \frac{\lambda }{4}$\\	
\hline
Noise power	& -80 dBm\\
\hline
Path loss exponents	& $\alpha  = 3.7,\tau  = 2.7$\\
\hline
Rician factors &	5\\
\hline
$P^d,P^u$ & 40 dBm\\
\hline
$P^d_{max},P^u_{max}$&20 dBm\\
\hline
Number of IoT nodes &	2 to 8\\
  \hline
\end{tabular}
\end{table}

\begin{figure}[!ht]
    \centering
    \centerline{\includegraphics[width=3.5in]{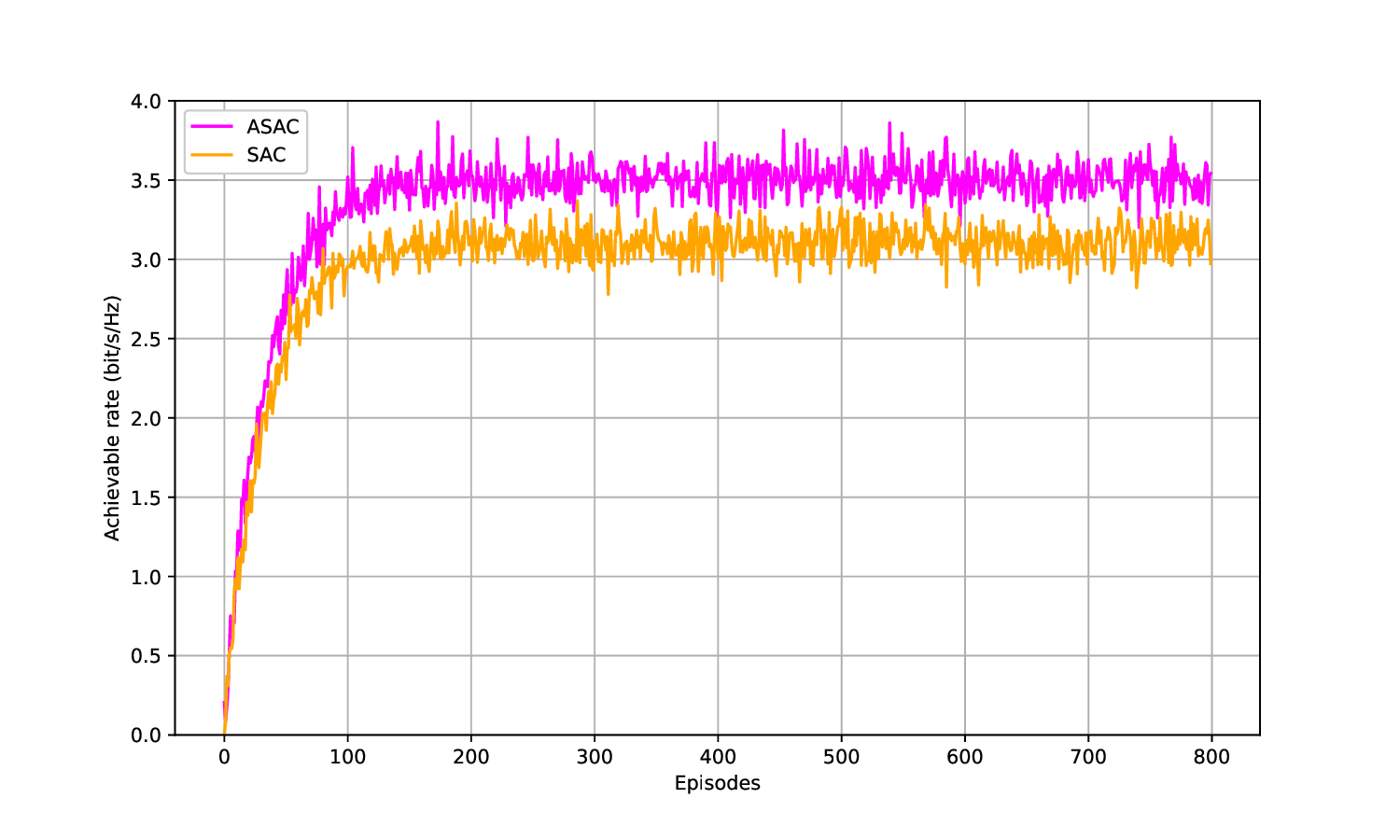}}
    \caption{Convergence of the proposed framework.}
    \label{fig:conv}
\end{figure}

Fig. \ref{fig:conv} depicts the evolution of achievable rate across training episodes for both the proposed ASAC algorithm and the SAC algorithm. As shown, the achievable rate under ASAC steadily increases over the training period, eventually stabilizing at a higher rate compared to SAC.
This difference is due to ASAC's adaptive mechanism, which enhances stability and exploration during training, allowing it to achieve a higher and more stable rate. SAC, on the other hand, lacks this adaptive approach, leading a lower overall achievable rate.

\begin{figure}[!ht]
    \centering
    \centerline{\includegraphics[width=3.5in]{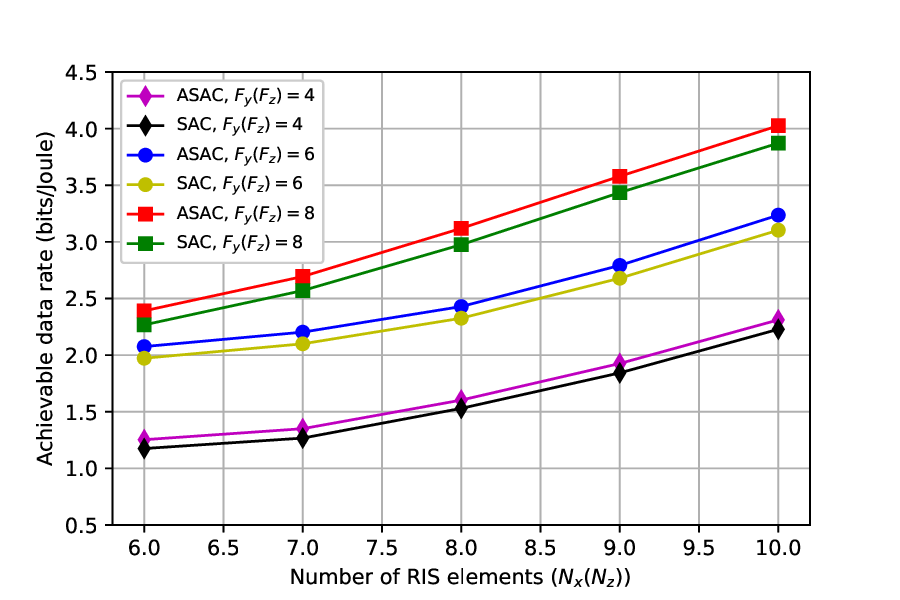}}
    \caption{Achievable rate for different number of RIS elements.}
    \label{fig:fig1}
\end{figure}

As illustrated in Fig. \ref{fig:fig1}, increasing the number of RIS elements leads to a higher achievable data rate. Focusing on this effect, Fig. \ref{fig:fig1} shows the impact of varying the number of ground RIS elements on the data rate while keeping the number of flying RIS elements, \( F \), constant. We assume both the flying RIS and ground RIS have square configurations, i.e., \( F_y = F_z \) and \( N_x = N_z \). It is evident that the achievable data rate increases with a higher number of RIS elements. The proposed ASAC method consistently outperforms SAC; for instance, when \( N_x (N_z) = 10 \), ASAC achieves an 8.22\% improvement over SAC.

\begin{figure}[!ht]
    \centering
    \centerline{\includegraphics[width=3.5in]{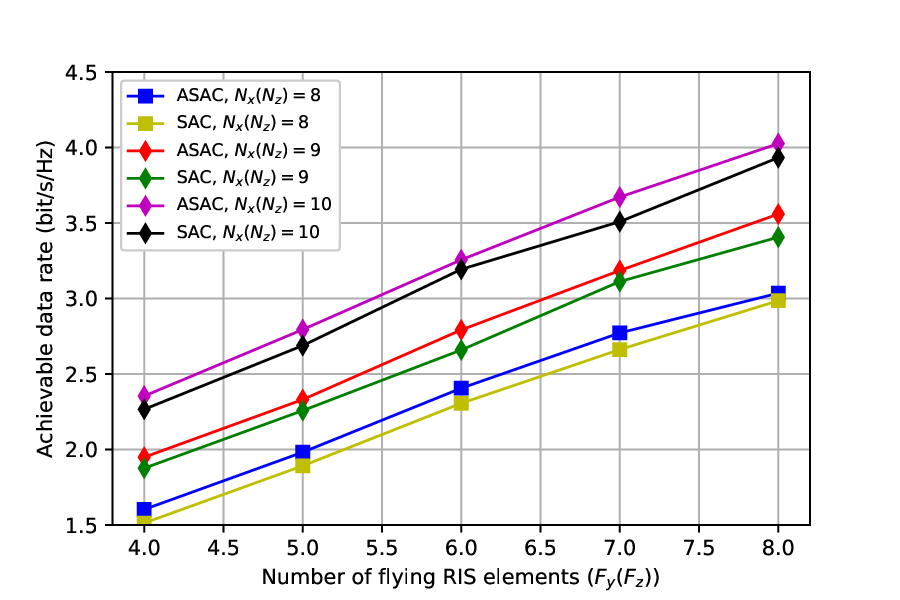}}
    \caption{Achievable rate for different number of flying RIS elements.}
    \label{fig:fig2}
\end{figure}
Fig. \ref{fig:fig2} demonstrates the case where the number of flying RIS elements is increased while the ground RIS element count remains fixed. In this scenario, the combination of flying RIS flexibility with the advantages of additional elements significantly enhances the achievable data rate. Although ASAC shows some fluctuations, its overall performance consistently exceeds that of SAC.

\begin{figure}[!ht]
    \centering
    \centerline{\includegraphics[width=3.5in]{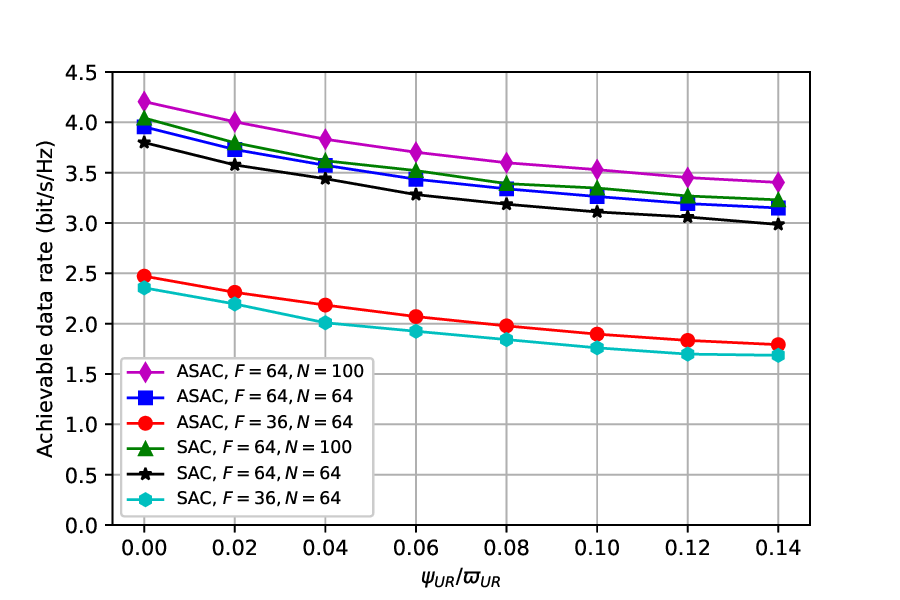}}
    \caption{Impact of UAV jittering on the achievable data rate.}
    \label{fig:fig3}
\end{figure}

The impact of the jittering for different sizes of flying RIS and ground RIS is illustrated in Fig. \ref{fig:fig3}. As we illustrated earlier during the modeling of the channel, the impact of the jittering leads to uncertainties in the elevation and the azimuth. We define the impact of the jittering as a ratio $\frac{\psi_{UR}}{\varpi_{UR}}$. From Fig. \ref{fig:fig3}, we can notice that the increasing in $\frac{\psi_{UR}}{\varpi_{UR}}$ (which means a drastic jittering) leads to decreasing data rate and that because more power is needed to compensate the impact of these jittering.  For larger RIS size, better achievable data rate can be obtained and that because less power is need to serve intended nodes since the signal can be easily directed toward the nodes. Again, the proposed ASAC clearly outperforms SAC.

\section{Conclusion}
In conclusion, this study demonstrates the potential of UAV and RIS for enhancing downlink communications in wireless networks. By jointly optimizing beamforming, phase shifts, and UAV trajectory, the proposed adaptive soft actor-critic (ASAC) framework effectively addresses the challenges of maximizing user fairness and coverage, even under conditions such as UAV jitter. Leveraging adaptive sparse transformers with attentive feature refinement (ASTAFER), the ASAC model dynamically adapts to real-time network conditions, providing a robust end-to-end solution that avoids traditional iterative or relaxation-based methods. Simulation results validate the superior performance of the ASAC-based approach over conventional SAC, underscoring its effectiveness and adaptability for real-time, fair, and efficient downlink communications in UAV-RIS networks. This framework offers promising directions for future wireless systems that require scalable and resilient solutions in dynamic environments.

\ifCLASSOPTIONcaptionsoff
  \newpage
\fi

\end{document}